# The ESSnuSB-plus (ESSnuSB+) Project: Status and Prospects

George Fanourakis
Institute of Nuclear & Particle Physics, NCSR Demokritos,
Agia Paraskevi, Attiki, Greece
On behalf of the ESSnuSB/ESSnuSB+ collaboration

**Abstract**

The ESSnuSB is a design study for a long-baseline neutrino experiment to precisely measure the CP violation in the leptonic sector, at the second neutrino oscillation maximum, using a beam driven by the uniquely powerful ESS linear accelerator. The ESSnuSB-plus (ESSnuSB+) design study program, which is an extension phase of the ESSnuSB project, aims in designing two new facilities, a Low Energy nuSTORM and a Low Energy Monitored Neutrino Beam and use them to precisely measure the (anti)neutrino-nucleus cross-section in the energy range of 0.2–0.6 GeV, where experimental data are lacking or imprecise. In addition, new target stations and a new water-Cherenkov near-near detector will be designed to measure cross sections and serve to explore sterile neutrino physics. An overall status of the project will be presented together with the ESSnuSB+ additions.

**Introduction – The ESSnuSB project**

The next generation neutrino experiments [1] are designed to resolve the question of neutrino mass ordering, to determine the quadrant of the atmospheric mixing angle $\theta_{23}$ via precision measurements and to prove the existence of CP violation (CPV). However, their precision in measuring the CP violating phase angle is not sufficient to resolve the theoretical models proposed to explain the lack of antimatter in the Universe [2,3]. ESSnuSB (ESS neutrino Super Beam) is a next-to-next generation long baseline (LBL) neutrino experiment under design, expected to follow up the LBL neutrino experiments currently in preparation, with the main purpose of measuring the CPV phase angle to such high precision as to enable the selection of the correct theory.

ESSnuSB proposes to use the powerful proton beam (5MW, 2.5 GeV proton kinetic energy) of the European Spallation Source (ESS) proton linear accelerator (linac), at Lund in southern Sweden, to create a high intensity neutrino super beam, with an energy distribution from 200 MeV/v to 600 MeV/c. The proposed LBL setup includes a near detector complex at ESS and a far detector complex at the Zinkgruvan mine of Sweden [4]. The combination of the neutrino beam energy with the location of the far detector places the experiment mostly on the second neutrino oscillation maximum, resulting in a dramatically improved matter-antimatter asymmetry largely immune to systematics and matter effects [4,5]. The Conceptual Design Report (CDR) of this study has been completed and published in 2022 [6]. The study intends to establish a parallel experimental facility at ESS, without interfering with ESS's main mission, i.e. to become the most powerful neutron facility in the world. The proton linac power is proposed to be doubled so to have equal power for the neutron and neutrino programs. The neutrino program requires the proton pulses to be shortened from 2.86 ms to 1.2 µs. This is accomplished by an accumulator which also requires that the linac accelerates $H^-$ instead of protons. A target station is designed with four targets and four horns to mitigate the effects of a high-power proton beam on a single target. A long baseline setup is proposed for the neutrino oscillation studies: A far water Cherenkov detector consisting of two tanks of 270 ktons each, placed in the Zinkgruvan mine, 360 km away, at a depth of about 1 Km, and a near detector complex consisting of a water Cherenkov, a super fine-grained scintillator detector and an emulsion detector. The CDR contains details of the ESS linac modifications and the performance of the detectors.

ESSnuSB follows the typical program of the LBL neutrino experiments, measuring the neutrino oscillation parameters, however, it especially focuses on the precision measurement of the CPV phase. Its ability to obtain high precision is due to the ESSnuSB (anti)neutrino spectrum mainly sitting at the second oscillation maximum. This fact makes the matter-antimatter asymmetry $A_{CPV}$ about 2.5 times larger than for the case of spectra at the first maximum [7], because of the larger interference term. This fact also makes the measurement of the CPV more stable against systematical errors. An important virtue of the ESSnuSB neutrino spectrum is also that the matter effects of neutrino interactions on their way to the far detector are much smaller at the second maximum (Fig. 1) thus significantly reducing the danger of faking CPV as when working in the first maximum [4].

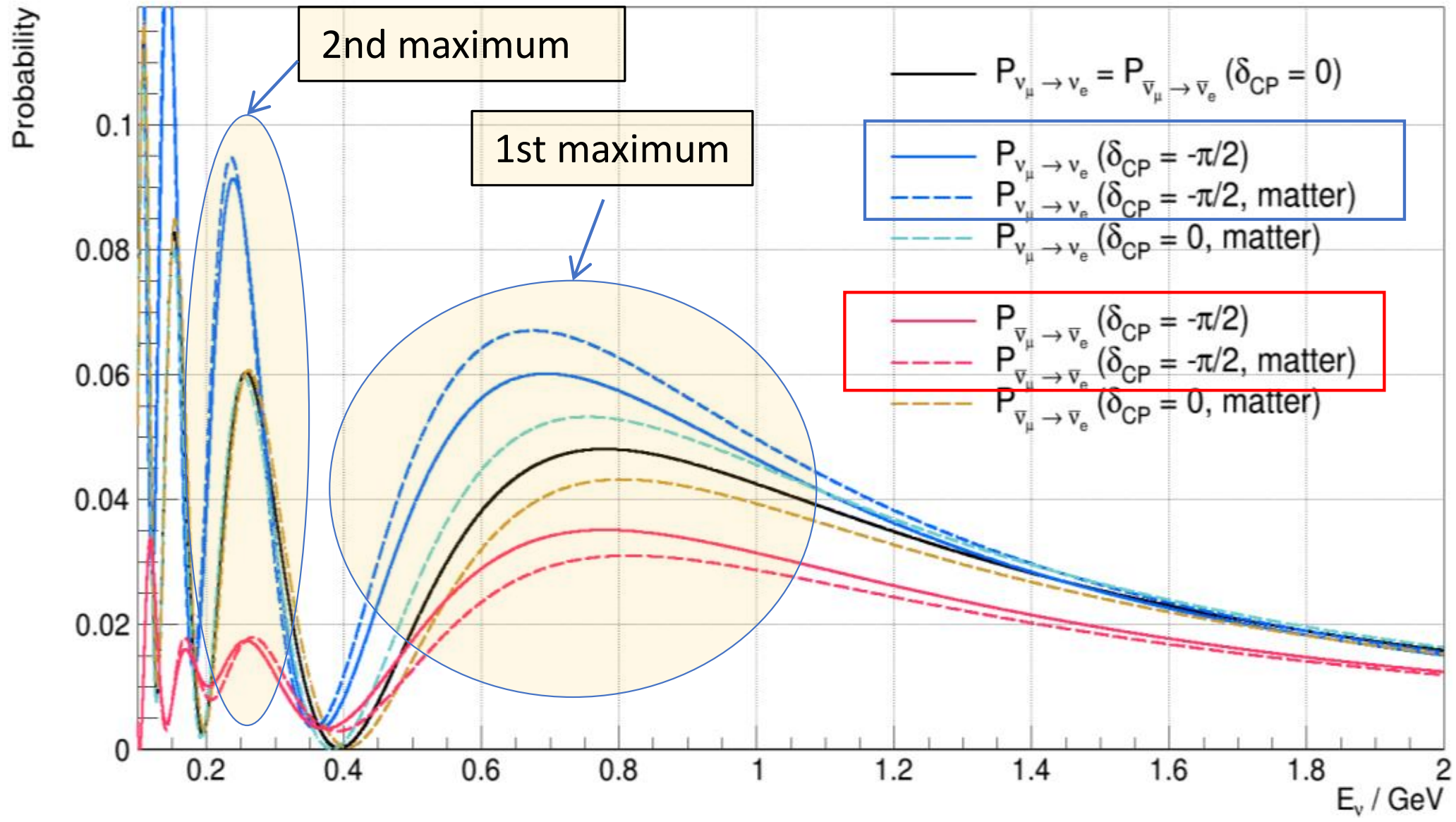

**Figure 1.** Neutrino and antineutrino disappearance oscillation probabilities as a function of neutrino energy at the fixed distance of 360 km. The oscillation probabilities are shown for $\delta_{CP}$ = 0 and $\delta_{CP}$ = $\pi$ /2. Full lines correspond to oscillations in vacuum and dashed lines to oscillations in matter.

The CPV violation measurement capability of ESSnuSB obtained from this study [6] is shown in Fig. 2. The results assume 5% normalization uncertainty and 10 years running shared equally between neutrinos and antineutrinos. The left panel shows the proposed experiment's sensitivity for CP-violation discovery as a function of $\delta_{CP}$ range of true values. It shows that at the maximal violation, i.e., $\delta_{CP} \sim \pm 90^o$, the CPV discovery sensitivity reaches about 12$\sigma$. The middle panel shows that, after10 years of data taking, ESSnuSB will cover

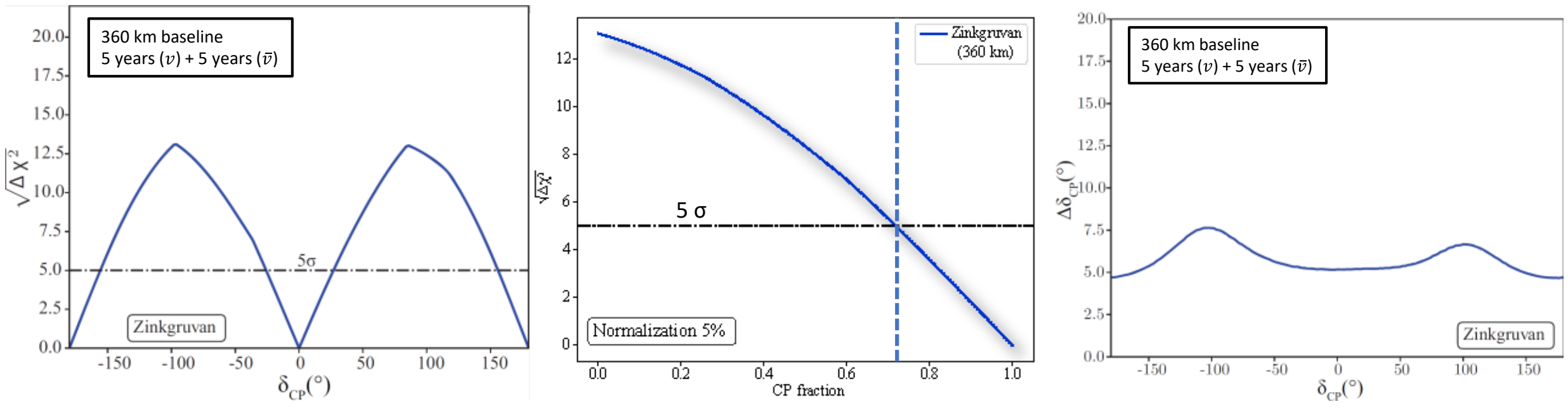

**Figure 2.** (Left panel) Discovery sensitivity of the leptonic CPV as a function of $\delta_{CP}$ range. (Middle) The fraction of the covered $\delta_{CP}$ range for which CPV can be discovered as a function of the significance. (Right) The precision on $\delta_{CP}$ value as a function of the $\delta_{CP}$ range.

more than 70% of the range of the true $\delta_{CP}$ values with a confidence level of more than 5$\sigma$ to reject the no-CPV

hypothesis. The right panel shows that the expected precision on the measured value of $\delta_{CP}$ will be better than $8^o$ for the all possible $\delta_{CP}$ parameters.

**The ESSnuSB+ (ESSnuSB-plus) project**

The uncertainty in the (anti)neutrino cross sections, especially below 600 MeV/c, is the dominant term of the systematic uncertainty of the ν-oscillation parameters measurements. Even though the effect of the systematics in the CP Violation measurement is much less in ESSnuSB it was deemed desirable to obtain new high precision results in this direction. Thus, after successfully completing the ESSnuSB design study, we have proposed the ESSnuSB+ project [8]. The purpose of ESSnuSB+ is to extend the Instrumentation infrastructure of ESSnuSB, by adding a) a Low Energy nuSTORM (LEnuSTORM), i.e. a muon racetrack storage ring, like the nuSTORM facility planned for CERN [9], b) a Low Energy Monitored neutrino beam (LEMNB) decay tunnel inspired by the ENUBET project [10] and c) a near-near detector for precise measurements of low energy (anti)neutrino cross section with water. Improvements are proposed as well to expand the Physics potential of ESSnuSB. The new project is being funded by the European commission Horizon-Europe program for the period 2023–2026. Fig. 3 shows the layout of the ESSnuSB/ESSnuSB+ accelerator and the near detector complex.

The design of two new target stations is foreseen for the new project, to generate pions which will feed the LEnuSTORM racetrack, and the LEMNB decay tunnel. These facilities will constitute the first phase of staging ESSnuSB, the main long baseline experiment. This phase will involve the precision neutrino cross section measurements with a new under design near-near water Cherenkov detector, the Low Energy neutrino from stored Muons and Monitored beam Near Detector (LEMMOND). The decay of negative polarity muons in the straight sections of LEnuSTORM will provide clean equal amounts of muon neutrinos and electron antineutrinos whereas the decay of positive muons will provide clean muon antineutrinos and electron neutrinos. LEMNB will also provide clean and precise fluxes of neutrinos and antineutrinos by tagging the muons from pion decays. These well-defined neutrino beams will be used with the LEMMOND detector to precisely measure low energy neutrino cross sections with water.

LEMMOND, the new detector added to the ESSnuSB detector suite, is a cylindrical Cherenkov with a fiducial volume of about 5 m diameter and 10 m length. It will be situated 50 m away from the new neutrino facilities LEnuSTORM or LEMNB.

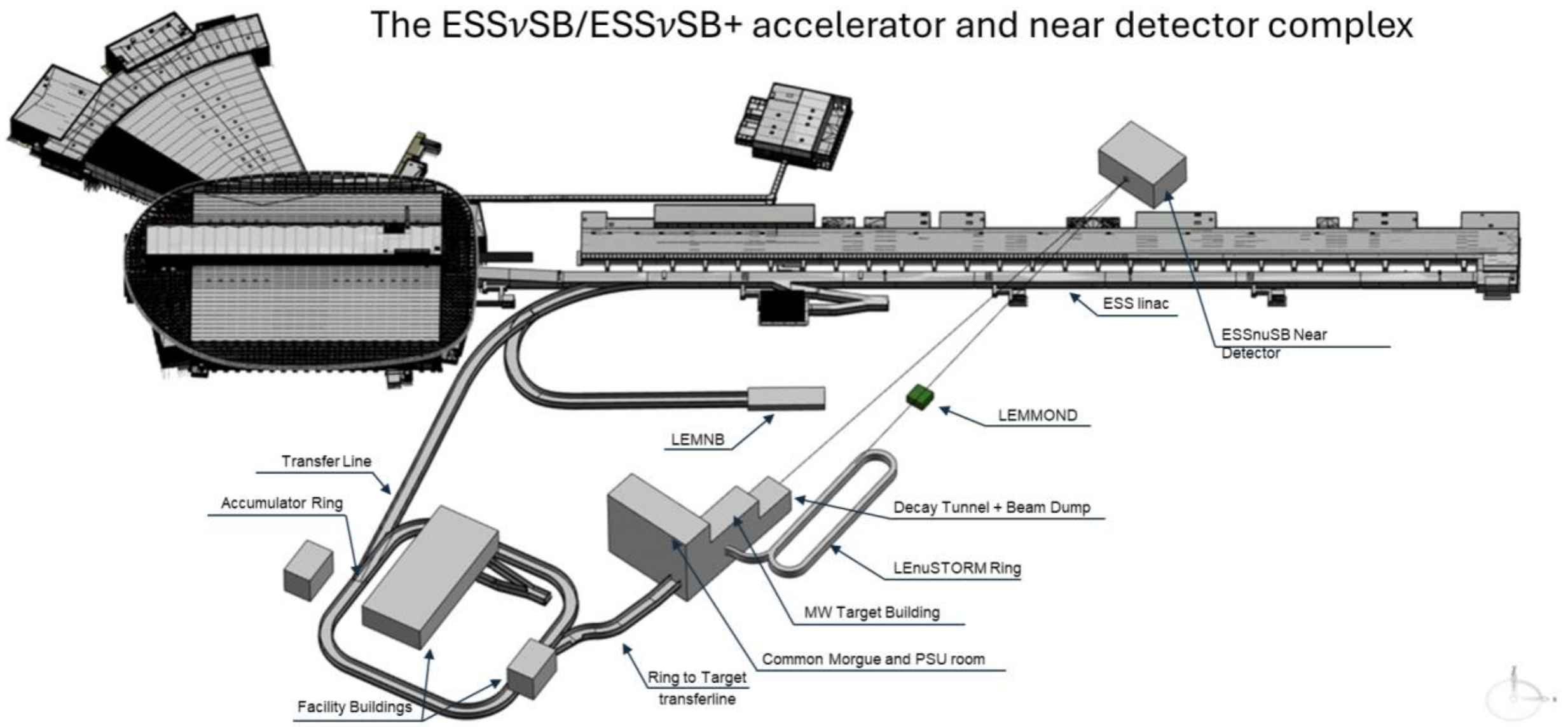

**Figure 3.** The layout of ESSnuSB and the additions of ESSnuSB+ at ESS.

An initial design of the simulation and reconstruction tools was tested for a simplified geometry of a flat detector plane made of 6400 5x5 $cm^2$ sensors. GEANT4 simulated muon tracks were generated with angles θ=0°, 30° and ϕ=0°, 200 cm away from the detector plane and their Cherenkov photons emission was followed to the sensors. A quantum efficiency of 25% and 25% coverage was assumed. Two choices for the time resolutions of the sensor elements were considered, 1.5 ns and 120 ps. The results of the analysis were quite encouraging; theta (θ) and phi (ϕ) angle resolutions of less than 1° were obtained for both timing capability choices. For the determination of the interaction point (the Z coordinate of the muon vertex) a precision of about 6 cm was obtained for the 1.5 ns sensors and less than 1 cm for the 120 ps sensors. These results were reproduced with full cylindrical geometry. Preliminary performance studies for the reconstruction of muons or electrons in the (anti)neutrino charged current (CC) interactions with water, using Graph Neural Networks (GNN) methodology show promising results. Fig. 4 displays the errors for the vertex position, the time of the interaction and the θ, ϕ angles for muons in the range of 200-600 MeV, using GNN.

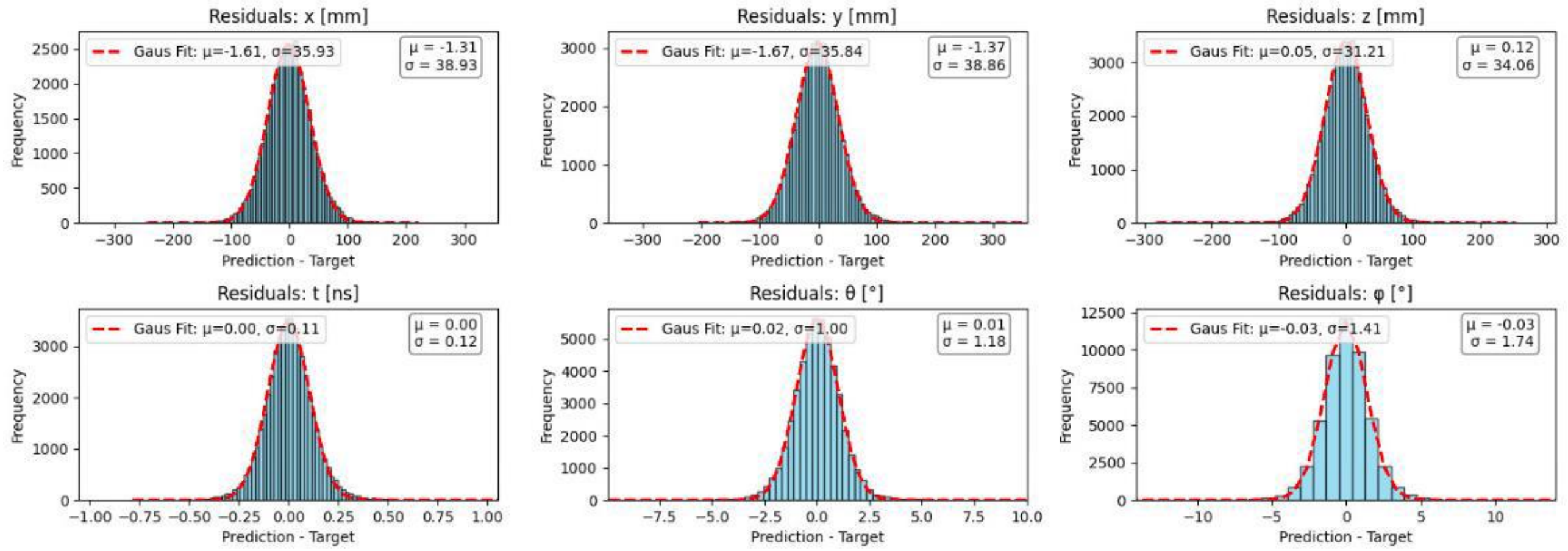

**Figure 4.** Parameter prediction errors for muon tracks of 200-600 MeV using GNN.

The LEnuSTORM and the combination of the LEMMOND detector with the near detector complex of ESSnuSB constitute a Short Base Line (SBL) setup able to investigate the existence of sterile neutrinos of mass square differences of 1 $eV^2$ to 10 $eV^2$.

Further analysis techniques and physics cases are being explored within the new project, besides the use of the new infrastructure for neutrino cross section measurements, the new option added of a SBL experiment for sterile neutrino investigations and the improvements considered for the detectors. The use of Graphic Neural Networks to improve the event classification of the ESSnuSB near Water Cherenkov detector has been completed [11]. A Monte Carlo study of neutrino oscillations using atmospheric neutrinos in the far detector was published [12] and showed that in 4 years ESSnuSB could determine the correct mass ordering with 3σ significance, it could determine the $\theta_{23}$ octant at 3σ in 4(7) years for normal (inverted) ordering and it could provide constraints on $\theta_{23}$ and $\Delta m^2_{31}$ (see Fig. 6). The addition of Gadolinium in the water Cherenkov detectors and the subsequent enhanced capability of final state neutron detection is currently also being investigated.

Work is progressing within the current project to investigate the sensitivity of ESSnuSB for Beyond the Standard Model (BSM) Physics. Results have been already published concerning the determination of constraints on scalar Non-Standard Interactions (NSI) parameters [13] and the determination of constraints on Quantum Decoherence parameters [14].

## Conclusions

ESSnuSB and its extension project ESSnuSB-plus involve the design of a next-to-next generation neutrino oscillation facility proposed. ESSnuSB main long base-line neutrino setup is dedicated to the precision measurement of the CP Violating phase. This setup includes a near detector suite made of a water Cherenkov, a Super fine-grained detector and an emulsion detector, and an underground far water Cherenkov detector.

ESSnuSB-plus is motivated by the lack of high precision (anti)neutrino cross sections with water at the ESSnuSB (anti)neutrino beam energy range and intends to precisely measure these cross sections. This will be accomplished by three new/novel facilities, comprising a LEnuSTORM ring, a LENMB instrumented neutrino decay tunnel and LEMMOND, a near-near water Cherenkov detector. The investigation of the use of these new facilities to probe the existence of sterile neutrinos with a short base-line setup is also a part of this project. More analysis techniques and physics capabilities are being evaluated during the ESSnuSB+ extension phase of the ESSnuSB project including GNN, atmospheric neutrinos, NSI and Quantum Decoherence.

**Acknowledgements**

This work is funded by the European Union. Views and opinions expressed are however those of the author(s) only and do not necessarily reflect those of the European Union. Neither the European Union nor the granting authority can be held responsible for them. We acknowledge further support provided by the following research funding agencies: Centre National de la Recherche Scientifique, France; Deutsche Forschungsgemeinschaft, Projektnummer 423761110 and the Excellence Strategy of the Federal Government and the Länder, Germany; Ministry of Science and Education of Republic of Croatia grant No. KK.01.1.1.01.0001; the European Union’s Horizon 2020 research and innovation programme under the Marie Skłodowska -Curie grant agreement No 860881-HIDDeN; the European Union NextGenerationEU, through the National Recovery and Resilience Plan of the Republic of Bulgaria, project No. BG-RRP-2.004-0008-C01.